\documentstyle[12pt]{article}
\newcommand{\beq}{\begin{equation}}
\newcommand{\eeq}{\end{equation}}
\newcommand{\beqa}{\begin{eqnarray}}
\newcommand{\eeqa}{\end{eqnarray}}
\newcommand{\beqan}{\begin{eqnarray*}}
\newcommand{\eeqan}{\end{eqnarray*}}
\newcommand{\ba}{\begin{array}}
\newcommand{\ea}{\end{array}}
\newcommand{\ben}{\begin{enumerate}}
\newcommand{\een}{\end{enumerate}}
\newcommand{\bfl}{\begin{flushleft}}
\newcommand{\efl}{\end{flushleft}}
\newcommand{\btab}{\begin{tabular}}
\newcommand{\etab}{\end{tabular}}
\newcommand{\bit}{\begin{itemize}}
\newcommand{\eit}{\end{itemize}}
\newcommand{\bdes}{\begin{description}}
\newcommand{\edes}{\end{description}}
\newcommand{\bdm}{\begin{displaymath}}
\newcommand{\edm}{\end{displaymath}}

\newcommand{\no}{\nonumber}

\newcommand{\ra}{\rightarrow}

\newcommand{\ve}{\varepsilon}

\newcommand{\dg}{\dagger}
\newcommand{\wt}{\widetilde}
\newcommand{\wh}{\widehat}

\newcommand{\cL}{{\cal L}}

\newcommand{\dfrac}{\displaystyle \frac}

\newcommand{\st}{\stackrel}
\newcommand{\gAz}{\stackrel{\circ}{g}_A}
\newcommand{\cor}{\ba{c} \wh{} \\[-14.5pt] - \ea}
\normalsize
\begin{document}
\begin{titlepage}
\begin{flushright}
UWThPh-1997-12\\
April 1997\\
\end{flushright}
\vspace{2.5cm}
\begin{center}

{\Large \bf Wave Function Renormalization in \\[10pt]
Heavy Baryon Chiral Perturbation Theory*}\\[40pt]
G. Ecker$^1$ and M. Moj\v zi\v s$^2$

\vspace{1cm}

${}^{1)}$ Institut f\"ur Theoretische Physik, Universit\"at Wien\\
Boltzmanngasse 5, A--1090 Wien, Austria \\[10pt]

${}^{2)}$ Department of Theoretical Physics, Comenius University\\
Mlynsk\'a dolina, SK--84215 Bratislava, Slovakia

\vfill
{\bf Abstract} \\
\end{center}
\noindent
We establish exact relations between relativistic form
factors and amplitudes for single--baryon processes and the corresponding
quantities calculated in the framework of heavy baryon chiral
perturbation theory. A crucial ingredient for the proper matching is the first
complete treatment of baryon wave function renormalization in heavy
baryon chiral perturbation theory.

\vfill
\noindent * Work supported in part by FWF (Austria), Project No. P09505--PHY,
by HCM, EEC--Contract No. CHRX--CT920026 (EURODA$\Phi$NE) and by
VEGA (Slovakia), grant No. 1/1323/96.
\end{titlepage}

\paragraph{1.}
Heavy baryon chiral perturbation theory (HBCHPT) \cite{JM91} is a method
to calculate Green functions and S--matrix elements for single--baryon
processes in a systematic chiral expansion. Although it singles out a special
reference frame characterized by a time--like unit four--vector $v$, Lorentz
invariance of the underlying meson--baryon Lagrangian \cite{GSS88,Kr90}
ensures that physical observables like S--matrix elements do not depend on
the choice of this frame \cite{LM92,EM96}.

The purpose of this letter is to derive the relativistic form factors and
amplitudes from the frame--dependent quantities of HBCHPT.
Our main result is that baryon wave function renormalization in HBCHPT
depends on the chosen frame through the baryon momentum. In coordinate space,
the wave function renormalization ``constant'' $Z_N$ is in fact
a non--trivial differential operator\footnote{A similar
situation arises in the nonrelativistic treatment of scalar field theories
\cite{GG97}.}.

\paragraph{2.}
We first recall the main ingredients of HBCHPT. For definiteness and because
the
chiral pion--nucleon Lagrangian has a well--established form up to $O(q^3)$
\cite{EM96}, we shall restrict the discussion to chiral $SU(2)$. With the
appropriate changes of coupling constants, the discussion and our results can
immediately be carried over to chiral $SU(3)$.

The starting point is the generating functional $Z[j,\eta,\bar\eta]$
of Green functions defined by the path integral \cite{GSS88}
\beq
e^{iZ[j,\eta,\bar\eta]} = N \int [du d\Psi d \bar \Psi]
\exp [i\{ \wt{S_M} + S_{MB} + \int d^4 x (\bar \eta \Psi + \bar \Psi \eta)\}]~.
\label{Z}
\eeq
The purely mesonic action $\wt{S_M}$ is the usual one \cite{GL84}
except that the low--energy constants are modified. The difference between
$S_M$ and $\wt{S_M}$ is due to the fermion determinant (closed nucleon
loops), analogous to the difference between the mesonic constants
for chiral $SU(2)$ and $SU(3)$, respectively. There, the
contributions of kaon and eta loops are included in the $SU(2)$ low--energy
constants just like the nucleon loops in the present case. The meson--baryon
action $S_{MB}$ corresponds to the relativistic pion--nucleon
Lagrangian \cite{GSS88}
\beq
\cL_{\pi N} = \bar \Psi (i \not\!\nabla - m + \dfrac{\st{\circ}{g}_A}{2}
\not\!u \gamma_5 ) \Psi + \ldots \label{piNrel}
\eeq
where $m,\st{\circ}{g}_A$ are the nucleon mass and the neutron decay constant
in the
chiral limit. The nucleon doublet is denoted as $\Psi$ and
$\eta,\bar \eta$ are fermionic sources. The covariant derivative $\nabla_\mu$
and the vielbein field $u_\mu$ are defined
as usual\footnote{Our notation is the same as in Refs.~\cite{EM96,Ec94b}.};
bosonic external fields are denoted collectively as $j$.

The functional  $Z[j,\eta,\bar\eta]$ generates fully relativistic
Green functions. In order to obtain a systematic chiral expansion,
one performs a frame--dependent decomposition of the nucleon field $\Psi$
in the functional integral (\ref{Z}). In this way,
the dependence on the nucleon mass $m$ is shifted from
the nucleon propagator to the vertices of an effective Lagrangian,
so that the integration over the new fermionic variables produces a systematic
low--energy expansion.

In terms of the velocity--dependent fields $N_v,H_v$ defined as \cite{Ge90}
\beqa
N_v(x) &=& \exp[i m v \cdot x] P_v^+ \Psi(x) \label{vdf} \\
H_v(x) &=& \exp[i m v \cdot x] P_v^- \Psi(x) \no
\eeqa
with
$$
P_v^\pm = \dfrac{1}{2} (1 \pm \not\!v)~, \qquad v^2 = 1 ~,
$$
the pion--nucleon action $S_{MB}$ takes the form
\beqa
S_{MB} &=& \int d^4 x \{ \bar N_v A N_v + \bar H_v B N_v +
\bar N_v B' H_v - \bar H_v C H_v\} \label{SMB} \\
A &=& iv \cdot \nabla + \st{\circ}{g}_A S \cdot u + \ldots \no \\
B &=& i \not\!\nabla^\perp - \dfrac{\st{\circ}{g}_A}{2} v \cdot u \gamma_5
 + \ldots, \qquad B'=\gamma^0 B^\dg \gamma^0 \no \\
C &=& 2m + i v \cdot \nabla + \st{\circ}{g}_A S \cdot u + \ldots,
\qquad \nabla_\mu^\perp = \nabla_\mu - v_\mu v \cdot \nabla   \no
\eeqa
with the spin matrix $S^\mu = i/2 \gamma_5 \sigma^{\mu\nu} v_\nu$.
Rewriting also the source terms in (\ref{Z}) in terms of
$N_v,H_v$ and corresponding sources
\beq
\rho_v = e^{imv \cdot x} P_v^+ \eta , \qquad
R_v = e^{imv \cdot x} P_v^- \eta ~, \label{src}
\eeq
one can integrate out the ``heavy'' components $H_v$
 to obtain a non--local action in the ``light''
fields $N_v$ \cite{MRR92,BKKM92,Ec94b}. By expanding $C^{-1}$ in a power series
in $1/m$,
\beq
C^{-1} = \dfrac{1}{2m} - \dfrac{i v \cdot \nabla + \st{\circ}{g}_A S \cdot
u}{(2m)^2}
+ \ldots ~,\label{cm1}
\eeq
this non--local action turns into a series of local actions of well--defined
chiral dimensions. Integration over $N_v$ leads to
\beq
e^{iZ[j,\eta,\bar\eta]} = N \int [du]
e^{i(S_M[u,j] + Z_{MB}[u,j,\eta,\bar\eta])} \label{mfi}
\eeq
where
\beqa
\lefteqn{Z_{MB}[u,j,\eta,\bar\eta] = - \int d^4 x \{ \bar \rho_v
(A + B' C^{-1} B)^{-1} \rho_v} \label{ZMBU} \\
&& \mbox{} + \bar R_v C^{-1} B(A + B' C^{-1} B)^{-1}
\rho_v + \bar \rho_v (A + B' C^{-1} B)^{-1} B' C^{-1} R_v \no \\
&& \mbox{} + \bar R_v C^{-1} B (A + B' C^{-1} B)^{-1}
 B' C^{-1} R_v - \bar R_v C^{-1} R_v\}.\no
\eeqa
Note that $S_M$ rather than $\wt{S_M}$ appears in (\ref{mfi}) due to an
interchange of limits: $C^{-1}$ has been expanded in (\ref{cm1}) before
functional integration over $N_v$. This makes the fermion determinant trivial
to any finite order in $1/m$ \cite{MRR92}.

{}From here on, the standard procedure of CHPT \cite{GL84}
can be applied: the action in the functional integral (\ref{mfi}) is
expanded around the classical solution $u_{\rm cl}[j]$ of the
lowest--order equation of motion. Integration over the quantum
fluctuations generates a systematic low--energy expansion for
$Z[j,\eta,\bar \eta]$. Although each term in this expansion of definite
chiral order depends on the chosen frame, the functional $Z[j,\eta,\bar \eta]$
is Lorentz invariant \cite{LM92,EM96} giving rise to fully
relativistic Green functions. Again due to the already mentioned interchange
of limits ($1/m$ expansion before functional or loop integrations), the
equivalence between Green functions calculated in the relativistic formalism
and in HBCHPT is strictly true only at tree level, but requires a proper
matching of low--energy constants beyond tree level\footnote{We are much
indebted to J\"urg Gasser for clarifying discussions on the relation between
relativistic and HBCHPT Green functions.}.

\paragraph{3.}
We have written (\ref{ZMBU}) on purpose as a functional of the original
fermionic
sources $\eta, \bar\eta$ to emphasize its relativistic character. The
decomposition
(\ref{src}), on the other hand, is frame dependent. It has been common
practice to neglect the ``heavy'' sources $R_v$ for processes involving
baryons (rather than anti--baryons). We will show that already to $O(q^3)$
this omission is not justified.

Although the functional $Z_{MB}[u,j,\eta,\bar\eta]$ still depends on the meson
fields $u(\phi)$, the nucleons have been integrated out. We can therefore
extract the structure of Green functions from this functional as far as
external nucleons are concerned. For non--trivial S--matrix elements, this
functional must
exhibit poles in momentum space in the in-- and outgoing nucleon momenta. Since
in HBCHPT $C^{-1}$ is expanded in a series of local operators, those poles can
only be due to the operator
$$
(A + B' C^{-1} B)^{-1}~.
$$
As a consequence, all terms in (\ref{ZMBU}) can contribute to S--matrix
elements
except for the contact term $\bar R_v C^{-1} R_v$. However, the three terms
involving the ``heavy'' source $R_v$ can only produce one--nucleon poles if
there are no external lines coming from
the factors $C^{-1}B$ or $B'C^{-1}$ next to $\bar R_v$ or  $R_v$.
Thus, the contributions of ``heavy'' sources appear only in external
nucleon propagators. For our subsequent
calculation of wave function renormalization to $O(q^3)$, only the
field--independent differential operators
\beqa
P^-_v C^{-1} B &=& \dfrac{1}{2m} P^-_v i \not\!\partial^\perp P^+_v + \ldots
\label{CB} \\
B' C^{-1} P^-_v &=& \dfrac{1}{2m} P^+_v i \not\!\partial^\perp P^-_v +
\ldots \no
\eeqa
will actually matter.

Let us first consider the two--point function.
{}From (\ref{mfi}) and (\ref{ZMBU}), after functional integration over the
meson fields, the nucleon propagator in momentum space has the general
form
\beq
S_N(p) = P^+_v S_{++}(k) P^+_v + P^+_v S_{+-}(k) P^-_v +
P^-_v S_{-+}(k) P^+_v + P^-_v S_{--}(k) P^-_v~, \label{SNp}
\eeq
with the off--shell momentum $p$ decomposed in the usual way as
$$
p = m v + k
$$
with a residual momentum $k$.

In the same way, we
deduce the structure of a general $n$--point function ($n \ge 3$):
\beqa
& &\left( P^+_v S_{++}(k_{\rm out}) P^+_v + P^-_v S_{-+}(k_{\rm out})
P^+_v\right) P^+_v T[j] P^+_v
\left(P^+_v S_{++}(k_{\rm in}) P^+_v + P^+_v S_{+-}(k_{\rm in}) P^-_v
\right) \no\\[5pt]
& & = S_N(p_{\rm out}) P^+_v T[j] P^+_v S_N(p_{\rm in})~,\label{GF}
\eeqa
with an obvious notation for the nucleon momenta. From here on, we always
neglect the contact term $\bar R_v C^{-1} R_v$ in (\ref{ZMBU}) because it
cannot contribute to S--matrix elements.
The functional $T[j]$ is the quantity that one calculates with the usual
HBCHPT Lagrangians \cite{JM91,EM96,BKKM92}, with the external nucleon
propagators removed. It depends on the bosonic external fields
$j$ and gives rise to Green functions and S--matrix elements along the
well--known rules of chiral perturbation theory \cite{GL84}. We emphasize
once again that $S_N(p)$ in (\ref{SNp}) and (\ref{GF}) is the fully
relativistic propagator.

\paragraph{4.}
We now turn to the calculation of the nucleon propagator to $O(q^3)$.
For this purpose, we recall the
pion--nucleon Lagrangian of HBCHPT in the formulation of Ref.~\cite{EM96}:
\beq
\wh \cL_{\pi N} =  \wh \cL_{\pi N}^{(1)} + \wh \cL_{\pi N}^{(2)} +
\wh \cL_{\pi N}^{(3)} + \ldots~,
\label{piNt}
\eeq
\beq
\wh \cL_{\pi N}^{(1)} = \bar N_v(iv \cdot \nabla + \st{\circ}{g}_A S \cdot u)
N_v~,
\label{piN1}
\eeq
\beqa
\label{piN2}
\wh \cL_{\pi N}^{(2)} &=& \bar N_v\left( - \dfrac{1}{2m} (\nabla \cdot
\nabla + i\st{\circ}{g}_A \{S \cdot \nabla, v \cdot u\}) \right.  \\
&& \mbox{} + \dfrac{a_1}{m} \langle u \cdot u\rangle +
\dfrac{a_2}{m} \langle (v \cdot u)^2\rangle +
\dfrac{a_3}{m} \langle \chi_+\rangle +
\dfrac{a_4}{m} \left( \chi_+ - \dfrac{1}{2} \langle \chi_+\rangle \right)
\no \\
&& \left. \mbox{} + \dfrac{1}{m} \ve^{\mu\nu\rho\sigma} v_\rho S_\sigma
[i a_5 u_\mu u_\nu + a_6 f_{+\mu\nu} + a_7 v_{\mu\nu}^{(s)}]\right) N_v \no
\eeqa
and $\wh \cL_{\pi N}^{(3)}$ can be found in Ref.~\cite{EM96}. The term
in (\ref{piN2}) with
coupling constant $a_3$ contributes to the nucleon mass:
\beq
 \langle \chi_+\rangle = 4 M^2 + \ldots ~,
\eeq
with $M$ the pion mass at lowest order, $O(q^2)$.

With the Lagrangian (\ref{piNt}), the nucleon propagator is given to $O(q^3)$
by
\beqa
S_{++}(k)^{-1} &=& v \cdot k + \dfrac{k^2}{2m} + \dfrac{4a_3 M^2}{m} -
\Sigma_{\rm loop} (v \cdot k) \label{nprop} \\
S_{+-}(k) &=& \dfrac{1}{2m} P^+_v S_{++} \not\!k^\perp P^-_v \no \\
S_{-+}(k) &=& \dfrac{1}{2m} P^-_v \not\!k^\perp S_{++} P^+_v \no~.
\eeqa
$S_{--}(k)$ does not contribute to $O(q^3)$.
The loop contribution of $O(q^3)$ to the nucleon self--energy is \cite{BKKM92}
\beqa
\Sigma_{\rm loop}(x) &=& - \dfrac{3\st{\circ}{g}_A^2}{(4 \pi F)^2}
\left\{ \dfrac{3}{4} x\left( M^2 - \dfrac{2}{3} x^2 \right)
\left( 32\pi^2 \Lambda(\mu) + \ln \dfrac{M^2}{\mu^2}\right) \right.
\label{Sigma} \\
&& \mbox{} + \left.(M^2 - x^2)^{3/2} \arccos \left( \dfrac{-x}{M} \right)
- \dfrac{x}{2} (M^2 - x^2) \right\} \qquad (x^2 < M^2) \no \\
\Lambda(\mu) &=& \dfrac{\mu^{d-4}}{(4\pi)^2} \left\{
\dfrac{1}{d-4} - \dfrac{1}{2} [ \ln 4\pi + 1 + \Gamma'(1)]\right\}~.\no
\eeqa
The nucleon pole is determined
entirely by $S_{++}(k)$ which can also be written as
\beq
S_{++}(k)^{-1} = \dfrac{1}{2m} \left\{ p^2 - m^2 + 8 a_3 M^2 - 2m
\Sigma_{\rm loop} (v \cdot k)\right\}~.\label{Spp}
\eeq
We define the on--shell nucleon momentum $p_N$ as
\beq
p = mv + k \st{\rm on-shell}{\longrightarrow} p_N = m_N v + Q~, \qquad
m_N = m + \Delta m~,
\eeq
\beq
 p_N^2=m_N^2 \quad \Longrightarrow \quad  2 m_N v \cdot Q + Q^2 =0~,
\eeq
where $Q$ is another residual momentum, $m_N$ is the physical nucleon
mass (in the isospin limit) and $\Delta m$ is at least $O(q^2)$.
Of course, we can always choose a frame for, say, the incoming nucleon with
$Q=0$.
But unless we are interested only in the forward direction, we can obviously
not
make the same choice for the outgoing nucleon as well.
On--shell we have
\beq
v \cdot k = \Delta m - \dfrac{Q^2}{2m_N} = O(q^2)~.
\label{vkon}
\eeq
Therefore, to $O(q^3)$ we find from (\ref{Sigma}), (\ref{Spp}) and (\ref{vkon})
\beq
m^2_N = m^2 - 8 a_3 M^2 + 2m \Sigma_{\rm loop}(0) + O(q^4)
\eeq
implying
\beq
\Delta m = - \dfrac{4a_3 M_\pi^2}{m_N} - \dfrac{3g^2_A \pi M_\pi^3}{2(4\pi
F_\pi)^2} + O(q^4) \label{Dm}
\eeq
in agreement with Refs.~\cite{GSS88,BKKM92}.

Wave function renormalization is more subtle. We
introduce yet another (arbitrary) four--vector $r$ to control the
on--shell limit $p\to p_N$ by letting the real parameter $\lambda$ tend to
zero:
\beqa
p &=& p_N + \lambda r  \\
k &=& \Delta m~ v + Q + \lambda r \no ~.
\eeqa
Although the final result for $Z_N$ must be
independent of how we approach the nucleon pole, the actual calculation to a
given
order will profit from a clever choice of $r$. Choosing $r=v$, we have
\beqa
v \cdot k &=& v \cdot p_N - m + \lambda \no \\
\Sigma_{\rm loop}(v \cdot k) &=& \Sigma_{\rm loop} (v \cdot p_N - m) +
\lambda \Sigma^\prime_{\rm loop}(v \cdot p_N - m) + O(\lambda^2)\\
\Sigma^\prime_{\rm loop}(v \cdot p_N - m) &=&\Sigma^\prime_{\rm loop}(0)+
O(q^3)\no
\eeqa
and thus
\beq
\Sigma^\prime_{\rm loop}(v \cdot p_N - m) = - \dfrac{9 g^2_A M_\pi^2}{2(4 \pi
F_\pi)^2}
\left[ (4\pi)^2 \Lambda(\mu) + \ln \dfrac{M_\pi}{\mu} + \dfrac{1}{3} \right]
+ O(q^3) \label{seprime}
\eeq
where the prime stands for the derivative.

Near the nucleon mass shell, we can write
\beqa
S_{++}(k)^{-1} &=& \dfrac{\lambda}{m} \left[ v \cdot p_N - m \Sigma^\prime_{\rm
loop}
(v \cdot p_N - m)\right] + O(\lambda^2)\label{Sppshell} \\
S_N(p) &=& \dfrac{m(P^+_v + \dfrac{1}{2m} \not\!k^\perp)}
{\lambda[v \cdot p_N - m \Sigma^\prime_{\rm loop} (v \cdot p_N - m)] +
O(\lambda^2)}~.
\eeqa
Referring to the general structure of Green functions in (\ref{GF}) and
recalling the relation between Green functions and S--matrix elements
for external fermions, we define the wave function renormalization ``constant''
$Z_N(Q)$ in the usual way as
\beq
Z_N(Q) u(p_N) = \lim_{p \ra p_N} S_N(p)(\not\!p - m_N) u(p_N)~,
\eeq
implying
\beq
Z_N(Q)u(p_N) = \dfrac{m(P^+_v + \dfrac{1}{2m} \not\!k^\perp) \not\!v
u(p_N)}{v \cdot p_N - m \Sigma^\prime_{\rm loop}(v \cdot p_N - m)}~.
\eeq
Neglecting consistently higher--order terms, we find
\beqa
Z_N(Q) &=& \dfrac{m}{m_N} \dfrac{v \cdot p_N - \dfrac{1}{2} v \cdot Q}
{v \cdot p_N - m \Sigma^\prime_{\rm loop}(v \cdot p_N - m)}  \\
&=& \dfrac{m}{m_N} \dfrac{1 + \dfrac{v \cdot Q}{2m_N}}
{1 + \dfrac{v \cdot Q}{m_N} - \Sigma^\prime_{\rm loop}(v \cdot p_N - m)} \no
\eeqa
and therefore finally
\beq
Z_N(Q) = 1 + \dfrac{4a_3 M_\pi^2}{m^2_N} + \dfrac{Q^2}{4m^2_N} -
\dfrac{9g^2_A M_\pi^2}{2(4\pi F_\pi)^2}
\left[(4\pi)^2 \Lambda(\mu) + \ln \dfrac{M_\pi}{\mu} + \dfrac{1}{3}\right]
+ O(q^3)~.\label{ZNQ}
\eeq

The wave function renormalization ``function'' $Z_N(Q)$ is our main result.
For $Q=0$, it agrees with the recent result of Fearing et al. \cite{FLMS97}.
However, as already emphasized, one cannot neglect the momentum dependence of
$Z_N(Q)$ altogether, except of course in the forward direction.

\paragraph{5.}
In principle, any choice of reference frame is equally acceptable due to
Lorentz
invariance of the theory. In practice, some choices will be more convenient
than
others for extracting amplitudes of a given order in the low--energy expansion.
An especially convenient choice is the ``initial--nucleon rest frame''
(INRF) defined as
\beqa
p_{\rm in} &=& m_N v = mv + k_1 \\
p_{\rm out} &=& m_N v + q = mv + k_2 \no \\
q &=& p_{\rm out} - p_{\rm in} = k_2 - k_1~, \qquad t=q^2 \no ~.
\eeqa
In other words, the INRF corresponds to
\beq
Q_{\rm in}=0~, \qquad Q_{\rm out}=q~.
\eeq
In this frame, wave function renormalization for single--nucleon
processes assumes the form
\beqa
\sqrt{Z_N^{\rm in} Z_N^{\rm out}} &=& \sqrt{Z_N(0) Z_N(q)}\label{Zinout} \\
&=& 1 + \dfrac{4a_3M_\pi^2}{m^2_N} +
\dfrac{t}{8m^2_N} - \dfrac{9 g^2_A M_\pi^2}{2(4 \pi F_\pi)^2}
\left[ (4\pi)^2 \Lambda(\mu) + \ln \dfrac{M_\pi}{\mu} + \dfrac{1}{3}\right]
+ O(q^3) ~.\no
\eeqa
The following relations are useful for actual calculations in the INRF:
\beqa
&& k^2_1 = \Delta m^2 = O(q^4), \qquad k^2_2 = \Delta m^2 +
\left( 1 - \dfrac{\Delta m}{m_N}\right) t = t + O(q^4)  \\
&& v \cdot k_1 = \Delta m, \qquad v \cdot k_2 = \Delta m - \dfrac{t}{2m_N},
\qquad k_1 \cdot k_2 = \Delta m^2 - \dfrac{\Delta m}{2m_N} t = O(q^4) ~.\no
\eeqa

\begin{table}[t]
\caption{Relations between relativistic covariants and the corresponding
quantities in the INRF with
$\bar u(p_{\rm out})\Gamma u(p_{\rm in}) = \bar u(p_{\rm out})P^+_v
\wh \Gamma P^+_v u(p_{\rm in})$.}
$$
\begin{tabular}{|c|c|} \hline
& \\
$\Gamma$ & $\wh \Gamma$ \\ \hline
& \\
$1$ & $1$ \\
& \\
$\gamma_5$ & $\dfrac{q \cdot S}{m_N(1 - t/4m^2_N)}$ \\
& \\
$\gamma^\mu$ & $\left( 1 - t/4m^2_N\right)^{-1}
\left( v^\mu + \dfrac{q^\mu}{2m_N} + \dfrac{i}{m_N} \ve^{\mu\nu\rho\sigma}
q_\nu v_\rho S_\sigma\right)$ \\
& \\
$\gamma^\mu \gamma_5$ & $2S^\mu - \dfrac{q \cdot S}{m_N(1 - t/4m^2_N)}
v^\mu$ \\
& \\
$\sigma^{\mu\nu}$ & $2 \ve^{\mu\nu\rho\sigma} v_\rho S_\sigma +
\dfrac{1}{2m_N(1 - t/4m^2_N)}
\{ i(q^\mu v^\nu - q^\nu v^\mu) + 2(v^\mu \ve^{\nu\lambda\rho\sigma}-
v^\nu \ve^{\mu\lambda\rho\sigma}) q_\lambda v_\rho S_\sigma\}$ \\
& \\ \hline
\end{tabular}\label{tab1}
$$
\end{table}
In the INRF, it is straightforward to derive
relations between HBCHPT amplitudes and their relativistic counterparts.
The results are collected in Table \ref{tab1}. Note that, in contrast
to (\ref{Zinout}), the relations in Table \ref{tab1} are exact, i.e., they
hold to all orders in the chiral expansion.

We can now summarize the procedure for obtaining relativistic S--matrix
elements
for a general one--nucleon process in HBCHPT. We concentrate on the fermionic
part because the bosonic part is well--known \cite{GL84}.
\bit
\item Calculate Green functions with the usual chiral Lagrangian of HBCHPT,
e.g.,
in the form of Ref.~\cite{EM96}. This amounts to considering only the term
$$
 \bar \rho_v (A + B' C^{-1} B)^{-1} \rho_v
$$
in the generating functional (\ref{ZMBU}). The relevant Green functions
are contained in the functional $T[j]$ in (\ref{GF}).
\item Amputate the external nucleon propagators and multiply with a factor
$\sqrt{Z_N^{\rm in} Z_N^{\rm out}}$ to account for nucleon wave function
renormalization. In the INRF to $O(q^3)$, the relevant expression is given
in (\ref{Zinout}).
\item Relate the HBCHPT amplitudes to the relativistic ones with the help
of Table \ref{tab1}. These relations are specific to the INRF.
\eit

\paragraph{6.}
To demonstrate the effect of wave function renormalization
for the matching between HBCHPT and relativistic
amplitudes, we investigate some examples at tree level. Thus,
for the following discussion we disregard all loop contributions
including the one in (\ref{Zinout}).

For the relativistic Lagrangian, we take the leading--order Lagrangian
(\ref{piNrel}) except for adding
a term of $O(q^2)$ to keep track of the nucleon mass correction proportional
to $a_3$:
\beq
\cL_{\rm rel} = \bar \Psi \left( i \not\!\nabla - m + \dfrac{\gAz}{2}
\not\!u \gamma_5 + \dfrac{a_3}{m} \langle \chi_+\rangle \right) \Psi~.
\label{Lrel}
\eeq
Our first example concerns the nucleon isovector vector form factors
$F_i^V(t)$ ($i=1,2$). From Table \ref{tab1}, we obtain the following exact
relation
between the relativistic and the HBCHPT matrix elements in the INRF:
\beqa
 \langle p_{\rm out}| \bar q \gamma^\mu \tau_a q| p_{\rm in}\rangle
&=& \bar u(p_{\rm out}) \tau_a [\gamma^\mu F^V_1(t) + \dfrac{i}{2m_N}
\sigma^{\mu\nu} q_\nu F^V_2(t)] u(p_{\rm in})  \label{FV12} \\
&=& \left( 1 - \dfrac{t}{4m^2_N}\right)^{-1} \bar u_+(p_{\rm out}) \tau_a
\left\{ \left[ F^V_1(t) + \dfrac{t}{4m^2_N} F^V_2(t)\right]
\left( v^\mu + \dfrac{q^\mu}{2m_N}\right) \right. \no \\
&& \mbox{} + \left. [F^V_1(t) + F^V_2(t)] \dfrac{i}{m_N}
\ve^{\mu\nu\rho\sigma} q_\nu v_\rho S_\sigma \right\} u_+(p_{\rm in})~,
\no
\eeqa
with
$$
u_+(p) = P^+_v \; u(p)~.
$$
The natural quantities emerging in the HBCHPT
calculation \cite{BKKM92} are the Sachs form factors
\beqa
G_E(t) &=& F^V_1(t) + \dfrac{t}{4m^2_N} F^V_2(t) \\
G_M(t) &=& F^V_1(t) + F^V_2(t) ~. \no
\eeqa

We concentrate here on the more illuminating case of $G_E(t)$.
To determine $G_E(t)$ from the respective Lagrangians, we trace the
external isovector vector field\footnote{To avoid confusion with
the unit vector $v$, we depart here from the standard notation for the external
vector field.} $V^\mu(x)$ in the covariant derivative
\beq
\nabla_\mu = \partial_\mu - iV_\mu \label{Vmu}+\ldots ~.
\eeq
The relativistic calculation with the Lagrangian (\ref{Lrel}) at tree level is
then trivial:
\beq
G_E(t) = 1~.\label{GErel}
\eeq
The HBCHPT calculation is not as trivial because even with the simple
Lagrangian
(\ref{Lrel}) the corresponding Lagrangian
\beq
\wh \cL_{\rm HBCHPT} = \bar N_v ( A + B' C^{-1} B ) N_v
\label{Lnonrel}
\eeq
consists of a whole tower of terms with increasing chiral dimensions due to the
expansion of $C^{-1}$. Since we have calculated wave function renormalization
at $O(q^3)$, we can check the equivalence with the relativistic
calculation up to the same order.

The relevant part of (\ref{Lnonrel}) for the calculation of $G_E(t)$
is\footnote{The last term in (\ref{LGE}) of $O(q^3)$ can be further decomposed
in the basis of Ref.~\cite{EM96}.}
\beq
\wh \cL_{\rm HBCHPT} = \bar N_v \left( iv \cdot \nabla
- \dfrac{1}{2m} \nabla \cdot \nabla - \dfrac{1}{8m^2}
[\nabla_\mu, [\nabla^\mu, iv \cdot \nabla]] \right) N_v + \ldots
\label{LGE}
\eeq
giving rise to the vertex
\beq
\left( 1 + \dfrac{t}{8m^2}\right) v \cdot V + \dfrac{1}{2m}(2 k_1
+ q) \cdot V \cor \left(1 + \dfrac{t}{8m^2} + \dfrac{\Delta m}{m}\right)
(v + \dfrac{q}{2m}) \cdot V ~.
\eeq
The difference between the two sides is of higher order. We have nicely
reproduced the relevant Lorentz structure for $G_E(t)$ in (\ref{FV12}) so that
we can immediately read off
\beq
\dfrac{G_E(t)}{ 1 - t/4m^2_N} =
\left(1 + \dfrac{t}{8m^2} + \dfrac{\Delta m}{m} \right)
\sqrt{Z_N^{\rm in} Z_N^{\rm out}}~.
\eeq
Inserting (\ref{Zinout}), we find indeed (to $O(q^3)$ in the matrix element)
\beq
G_E(t) =  1
\eeq
in agreement with the relativistic result (\ref{GErel}).

Another instructive example is the isovector axial form factor $G_A(t)$
defined through the matrix element (neglecting second--class currents)
for the isovector axial current
\beqa
\langle p_{\rm out}| \bar q \gamma^\mu \gamma_5 \tau_a q| p_{\rm in}
\rangle &=&
\bar u(p_{\rm out}) \tau_a \left[ \gamma^\mu \gamma_5 G_A(t) +
\frac{q^\mu}{2m_N} \gamma_5 G_P(t)\right] u(p_{\rm in})\label{GAP}  \\
&=& \left( 1 - \frac{t}{4m^2_N}\right)^{-1} \bar u_+(p_{\rm out}) \tau_a
\left\{ \left[ 2 \left(1 - \frac{t}{4m^2_N}\right) S^\mu -
\frac{q \cdot S}{m_N} v^\mu \right] G_A(t) \right. \no \\
&& \left. \mbox{} + \frac{q \cdot S}{2m^2_N} q^\mu G_P(t)\right\}
u_+(p_{\rm in})~. \no
\eeqa
To extract $G_A(t)$ from the Lagrangians (\ref{Lrel}) and (\ref{Lnonrel}),
we trace the external isovector axial--vector field $a^\mu(x)$ in the
vielbein field
\beq
u^\mu = 2 a^\mu +\ldots
\eeq
The relativistic calculation with (\ref{Lrel}) is again trivial:
\beq
G_A(t) = \gAz ~.\label{GArel}
\eeq
The HBCHPT calculation is more involved than in the previous case because
one has to account for the field transformations used in Ref.~\cite{EM96} to
eliminate equation--of--motion terms. We leave it as an exercise to verify
that, after some algebra, the relevant piece of the Lagrangian
(\ref{Lnonrel}) is of the form
\beq
\wh \cL_{\rm HBCHPT}
\cor \bar N_v \left\{ \gAz S \cdot u \left( 1 + \dfrac{\Delta m}{m}\right)
- \dfrac{i \gAz}{2m} \{S \cdot \nabla, v \cdot u\}
+ \dfrac{\gAz}{8m^2} [\nabla_\mu,[\nabla^\mu, S \cdot u]] \right\} N_v~.
\label{Laxial}
\eeq
To $O(q^3)$, this Lagrangian produces of course the appropriate
Lorentz structure for $G_A(t)$ in (\ref{GAP}). From (\ref{GAP}) and
(\ref{Laxial}) we infer
\beq
G_A(t) = \gAz \left( 1 + \dfrac{\Delta m}{m} - \dfrac{t}{8m^2} \right)
\sqrt{Z_N^{\rm in} Z_N^{\rm out}}
\eeq
and thus to $O(q^3)$
\beq
G_A(t) =  \gAz ~,
\eeq
again in agreement with the relativistic result (\ref{GArel}).

Finally, we list the relevant relations for elastic pion--nucleon scattering.
With the kinematics defined by
\beq
\pi(q_1) + N(p_{\rm in}) \ra \pi(q_2) + N(p_{\rm out})~,
\eeq
the usual definition of invariant amplitudes is (ignoring isospin)
\beq
\bar u(p_{\rm out})[A(\nu,t) + \not\!q_1 B(\nu,t)] u(p_{\rm in}) =
 \bar u(p_{\rm out}) \left[D(\nu,t) + \frac{i}{2m_N} \sigma^{\mu\nu}
q_{2\mu} q_{1\nu} B(\nu,t) \right] u(p_{\rm in})
\eeq
$$
D = A + \nu B, \qquad \nu = \frac{s - u}{4m_N} =
\frac{q_1 (p_{\rm in} + p_{\rm out})}{2m_N}~.
$$
In HBCHPT, the same matrix element can be decomposed as
\beq
\bar u_+(p_{\rm out}) \left[\alpha(\nu,t) + i \ve^{\mu\nu\rho\sigma}
q_{1\mu} q_{2\nu} v_\rho S_\sigma \beta(\nu,t) \right]
u_+(p_{\rm in})~.
\eeq
Table \ref{tab1} yields the translation between relativistic
and HBCHPT amplitudes in the INRF:
\beqa
A(\nu,t) &=& \alpha(\nu,t) + m_N \nu \beta(\nu,t)  \\
B(\nu,t) &=& - m_N \left(1 - \dfrac{t}{4m^2_N} \right) \beta(\nu,t)~.\no
\eeqa
The complete calculation of $\alpha$ and $\beta$ to $O(q^3)$,
including the proper wave function renormalization (\ref{Zinout}), can be
found in Ref.~\cite{MM97}.

\paragraph{7.}
Most calculations in HBCHPT for chiral $SU(2)$ have not been done with our
favourite Lagrangian in Ref.~\cite{EM96}, but with the original version of
Ref.~\cite{BKKM92} that differs by equation--of--motion terms.
Wave function renormalization depends on the form of the
Lagrangian even though S--matrix elements are unchanged. It is therefore
worthwhile to repeat the previous discussion of the nucleon propagator with the
Lagrangian of Ref.~\cite{BKKM92}.

Since the loop contribution is unchanged, we suppress it for the time
being and reinsert it only in the final result for $Z_N(Q)$. In the present
framework, the only difference to the previous case is in the function
\beq
S_{++}(k)^{-1} = v \cdot k + \dfrac{1}{2m} [k^2 - (v \cdot k)^2] +
\frac{4a_3 M^2}{m} + \mbox{loops} ~.\label{Sppold}
\eeq
Comparing with (\ref{nprop}), we observe that the equation--of--motion terms in
the second--order Lagrangian of Ref.~\cite{BKKM92} induce a term proportional
to $(v \cdot k)^2$. Even though (\ref{Sppold}) is not of the simple form
(\ref{Spp}), the pole position is of course unchanged, giving the same mass
correction (\ref{Dm}) as before. On the other hand, wave function
renormalization
is simpler here because instead of (\ref{Sppshell}) one now has
\beq
S_{++}(k)^{-1} = \lambda + O(\lambda^2)
\eeq
near the mass shell leading to
\beq
Z_N(Q) = \dfrac{1}{m_N} \left(v \cdot p_N - \dfrac{v \cdot Q}{2}\right)
= 1 + \frac{v \cdot Q}{2m_N} = 1 - \frac{Q^2}{4m^2_N}
\eeq
to the required accuracy. Reinserting the loop contribution, we compare the
final
result with the previous one in Table \ref{tab2}. As expected, the two
functions
$Z_N(Q)$ differ due to the field transformation necessary to pass from one
form of the Lagrangian to the other. Since that field transformation is
accomplished by a non--trivial differential operator \cite{EM96}, the change
of sign in the term proportional to $Q^2$ should not come as a surprise.

\begin{table}[t]
\caption{Wave function renormalization $Z_N(Q)$ for the Lagrangians of
Refs.~\protect\cite{EM96} and \protect\cite{BKKM92}, respectively, with
$\Sigma^\prime_{\rm loop} = - \dfrac{9 g^2_A M_\pi^2}{2(4\pi F_\pi)^2}
\left[ (4\pi)^2 \Lambda(\mu) + \ln \dfrac{M_\pi}{\mu} + \dfrac{1}{3}
\right]$.}
\vspace*{.4cm}
$$
\begin{tabular}{|c|c|} \hline
Lagrangian & $Z_N(Q)$ \\ \hline
& \\
EM \protect\cite{EM96} &
$1 + \dfrac{Q^2}{4m^2_N} + \dfrac{4a_3 M_\pi^2}{m_N^2} +
\Sigma^\prime_{\rm loop} + O(q^3)$ \\
& \\
BKKM \protect\cite{BKKM92} &
$ 1 - \dfrac{Q^2}{4m^2_N} + \Sigma^\prime_{\rm loop} + O(q^3)$ \\
& \\ \hline
\end{tabular}\label{tab2}
$$
\end{table}

\paragraph{8.}
We have analysed the structure of the generating functional of Green functions
in HBCHPT. This analysis has led to the first complete treatment of baryon
wave function renormalization to $O(q^3)$. We have shown that wave function
renormalization in HBCHPT cannot be described by a constant. Instead, in
momentum space it depends on the chosen frame via the baryon momentum.

We have also presented the general relations between HBCHPT and relativistic
amplitudes in the convenient initial--nucleon rest frame. Those relations are
exact, i.e., they hold to all orders in the low--energy expansion. We have
checked
the correspondence for the vector and axial--vector nucleon form factors at
tree
level. The correct expression for wave function renormalization is
crucial for this correspondence.

With our results, the relativistic amplitudes for any single--baryon process
can unambiguously be calculated in HBCHPT to $O(q^3)$, e.g., for elastic
pion--nucleon scattering \cite{MM97}.

\vfill

\section*{Acknowledgements}
\noindent We are grateful to J. Gasser for clarifying discussions on
the relation between the relativistic and the HBCHPT framework and for
helpful comments on the manuscript. We also thank G. M\"uller and S.
Steininger for discussions and comments.

\newpage

\newcommand{\PL}[3]{{Phys. Lett.}        {#1} {(19#2)} {#3}}
\newcommand{\PRL}[3]{{Phys. Rev. Lett.} {#1} {(19#2)} {#3}}
\newcommand{\PR}[3]{{Phys. Rev.}        {#1} {(19#2)} {#3}}
\newcommand{\NP}[3]{{Nucl. Phys.}        {#1} {(19#2)} {#3}}


\begin{thebibliography}{99}
\bibitem{JM91}
E. Jenkins and A.V. Manohar, \PL{B255}{91}{558}.
\bibitem{GSS88}
J. Gasser, M.E. Sainio and A. \v Svarc, \NP{B307}{88}{779}.
\bibitem{Kr90}
A. Krause, Helvetica Phys. Acta 63 (1990) 3.
\bibitem{LM92}
M. Luke and A.V. Manohar, \PL{B286}{92}{348};\\
R. Sundrum, Reparameterization invariance to all orders in heavy quark
effective theory, BUHEP-97-14, hep-ph/9704256, and references therein.
\bibitem{EM96}
G. Ecker and M. Moj\v zi\v s, \PL{B365}{96}{312}.
\bibitem{GG97}
A. Gall and J. Gasser, private communication.
\bibitem{GL84}
J. Gasser and H. Leutwyler, Ann. Phys. 158 (1984) 142; \NP{B250}{85}{465}.
\bibitem{Ec94b}
G. Ecker, \PL{B336}{94}{508}.
\bibitem{Ge90}
H. Georgi, \PL{B240}{90}{447}.
\bibitem{MRR92}
T. Mannel, W. Roberts and Z. Ryzak, \NP{B368}{92}{204}.
\bibitem{BKKM92}
V. Bernard, N. Kaiser, J. Kambor and U.-G. Mei\ss ner, \NP{B388}{92}{315}.
\bibitem{FLMS97}
H.W. Fearing, R. Lewis, N. Mobed and S. Scherer, Muon capture by a proton in
heavy baryon chiral perturbation theory, TRI-PP-97-5, MKPH-T-97-7,
hep-ph/9702394, to appear in Phys. Rev. D.
\bibitem{MM97}
M. Moj\v zi\v s, Elastic $\pi N$ scattering to $O(p^3)$ in heavy baryon chiral
perturbation theory, Univ. Bratislava preprint, hep-ph/9704415, to appear in
Z. Phys. C.


\end{thebibliography}
\end{document}